\documentclass[aps,pre,onecolumn,showpacs,showkeys,a4paper]{revtex4}
\input epsf  
\usepackage{graphicx} 
\usepackage{amsmath}
\usepackage{amssymb}
\usepackage{amscd}

\begin{document} 

\title{Stochastic bifurcations: a perturbative study}
\author{S\'ebastien Auma\^ \i tre$^1$, Kirone Mallick$^2$, 
Fran\c cois P\'etr\'elis$^1$
}
\affiliation{$^1$Laboratoire de Physique Statistique, 
Ecole Normale Sup\'erieure, 24
rue Lhomond, 75005 Paris, France\\
$^2$Service de Physique Th\'eorique, Centre d'\'etude de Saclay,91191 
Gif--sur--Yvette Cedex, France} 

\begin{abstract}

  We study a noise-induced  bifurcation  in the vicinity of the
 threshold  by using a perturbative expansion of the order parameter,
 called the Poincar\'e-Lindstedt expansion.  Each term of this series
 becomes divergent in the long time limit if the power spectrum of the
 noise does not vanish at zero frequency. These divergencies have a
 physical consequence~: they modify  the scaling of all the moments of
 the  order parameter near the threshold and lead to a multifractal
 behaviour.  We derive  this anomalous  scaling behaviour
 analytically   by a  resummation of  the  Poincar\'e-Lindstedt
 series and show that  the usual,  deterministic,  scalings are
 recovered  when the noise has a low frequency cut-off.   Our analysis
 reconciles  apparently  contradictory  results   found in the
 literature.
 \end{abstract}


\maketitle

\section{Introduction} 

  A random noise can  strongly   affect  the qualitative  behaviour 
 of a  nonlinear dynamical system  by  shifting the  bifurcation threshold,
 by  modifying  the characteristic exponents or   by  inducing  unexpected
 transitions  \cite{lefever,anishchenko}.  A  straightforward  approach 
 to study  the effect of   noise
 on a bifurcation diagram    would consist in  analyzing 
  the  nonlinear  Langevin   equation
 that governs the system.  However,    the interplay
 between noise and nonlinearity  results  in subtle effects
 that make  nonlinear stochastic 
 differential equations hard to   handle \cite{vankampen,gardiner}.   
  Therefore,  most  of the works   either consider    the
 linearized stochastic system in the neighbourhood of a stable
 manifold  (the nonlinearity is thus  eliminated)
  or analyze   Fokker-Planck type
  evolution equations for the Probability Distribution 
 Function (PDF)
 (the noise is  thus integrated out  by mapping  a
 stochastic  ordinary  differential  equation    into
 a deterministic partial differential  equation 
   in the phase space of the system).  Any  type  of noise can
  be studied by  the    linearization technique,  
  however  the   behaviour
 of the  linearized dynamics   is in general 
   different from that of  the
 real   nonlinear   system. On the other hand,  
  Fokker-Planck equations  fully take into account  the  nonlinear
 dynamics but   are valid  only  for 
 white noise (the various `Effective    Fokker-Planck Equations'  that 
 have been proposed to treat  colored noise  
 have a restricted range of validity).  
 The  Fokker-Planck  approach   can not therefore  be  applied to 
  a noise with an arbitrary spectrum.

The study of  a  nonlinear  system   
 subject to a deterministic forcing  
 is a  classical problem in  the theory of dynamical systems
 (including,  for example,  the phenomenon of parametric
 resonance of an oscillator driven by a  periodic modulation). 
 Many mathematical methods have been developed to analyze this question, 
 amongst  them  the  Poincar\'e-Lindstedt expansion  which is 
   a systematic  perturbative calculation free of secular
 divergences \cite{drazin,kevorkian}. 
 It is  natural   to try to adapt the  
 Poincar\'e-Lindstedt expansion   to the case of 
 nonlinear  system   driven by a noise \cite{luecke1,luecke2}:  such an 
  expansion is  versatile enough 
  to deal with  a {\it nonlinear}  system subject to 
 a noise having  {\it an arbitrary spectrum}, thus  allowing 
 one to  study precisely the impact of the spectral properties
 of the noise on a stochastic bifurcation.

 The aim of the present  work is to apply the  Poincar\'e-Lindstedt
  technique  to the study of a stochastic Hopf bifurcation  and 
 to determine the scaling exponents in the vicinity
 of the bifurcation threshold. We shall show that
  the characteristics of the noise have a strong influence
 on the scaling behaviour.  In particular the existence
 of low frequencies in the noise power spectrum results in 
 multifractality.  The perturbative expansion also  explains 
  the crossover between normal scaling and anomalous  scaling 
 and allows to resolve some controversial claims  in the literature.  
  
 The plan of this work is as follows:  
 In section \ref{sec:Poincare}, we define the model under study,  
   carry out   the  Poincar\'e-Lindstedt expansion and find 
 the conditions under which  this expansion becomes divergent.  
  In section  \ref{sec:resumation},
 we extract the leading term of this expansion to all orders
 and show that    after resummation  the normal scaling behaviour is 
 replaced by   multifractality. Concluding remarks are given in the 
  last section.

 \section{Perturbative analysis of a noisy Hopf bifurcation}
\label{sec:Poincare}

 The random dynamical system    we  study  here 
 has been widely  discussed  as a paradigm for a noise-induced
 bifurcation  \cite{lefever,graham} and is gouverned by the
 following stochastic equation~:
\begin{equation}
    \dot{x} =
  \left( \epsilon   + \Delta \xi(t) \right) x -  x^{2p+1} \, , 
\label{eq:Hopf}
\end{equation}
 where  $p$  is a strictly positive integer  and the noise $\xi(t)$ is
  a Gaussian  stationary random process  with  zero mean value 
  and with a correlation function given by
\begin{equation}
  \langle  \xi(t)  \xi(t')  \rangle  = {\mathcal D}(|t - t'|) \, .
\label{eq:corr}
\end{equation}
The power spectrum of the noise is the Fourier
 transform of the correlation function
\begin{equation}
  \hat{\mathcal D}(\omega) =  \int_{-\infty}^{+\infty} {\rm d}t \exp(i\omega t)
    \langle  \xi(t)  \xi(0)  \rangle  = 
  \int_{-\infty}^{+\infty} {\rm d}t \exp(i\omega t)
{\mathcal D}(|t|) \, .
\label{eq:PSD}
\end{equation}
When the noise   $\xi(t)$ is  white, the correlation function 
 ${\mathcal D}$ is a Dirac delta function and the power spectrum
    is a constant. We also recall that,
 because of the Wiener-Khinchin theorem \cite{vankampen}, the function
 $\hat{\mathcal D}(\omega)$  is non-negative. 

 Applying  elementary  dimensional analysis to   equation~(\ref{eq:Hopf}),
 we obtain  the  following scaling relations:
 \begin{equation}
  x \sim t^{{1}/{(2p)}} \, , \,\,\,\,\,\,\,\,
 \xi \sim t^{-1/2}  \, ,  \,\,\,\,\,\,\,\,
  \epsilon \sim \Delta^2 \sim t^{-1 } \, .
\label{eq:dimensions}
\end{equation}
The dimension of the noise  $\xi$ is so  chosen as to render the   
 power spectrum  $\hat{\mathcal D}(\omega)$  a dimensionless function. 

\subsection{The white noise case}

 When $\xi(t)$ is a Gaussian white noise, the  stationary
 solution  of the   Fokker-Planck equation
  corresponding to equation~(\ref{eq:Hopf}) is given by
  \begin{equation}
    P_{{\rm stat}}(x)  =  \frac{ 2p} 
  {\Gamma( \alpha)  (p\Delta^2)^{\alpha} } 
  x^{2p\alpha  -1} 
  \exp\left(-\frac{x^{2p}}{p\Delta^2} \right)   \,\,\,  \,\,\,
{\rm  with } \,\,\,   \alpha =    \frac{\epsilon}{p\Delta^2}   \, ,  
\label{eq:PDFblanc}
\end{equation}
 and  where $\Gamma$ represents the Euler Gamma-function.
 The bifurcation  threshold  is given by  $\epsilon = 0$; 
  for $\epsilon < 0 $, the solution~(\ref{eq:PDFblanc}) is not
  normalizable~:   the stationary PDF is the  Dirac distribution $\delta(x)$
 localized at the absorbing   fixed point  $x =0$. 
 For  $\epsilon >  0 $, the solution~(\ref{eq:PDFblanc}) is
 normalizable and is the extended   
 stationary PDF.  In this  case,  the moments of $x$  are given by 
\begin{equation}
   \langle x^{2n} \rangle  =    \frac{ \Gamma( \alpha + n/p) } 
  {\Gamma( \alpha) }   (p\Delta^2)^{n/p}  \, .  
\label{eq:momentblanc}
\end{equation}
 In the vicinity of the threshold,  $\epsilon$ is small and 
 we find that the moments scale linearly with  $\epsilon$, {\it i.e.},
 \begin{equation}
   \langle x^{2n} \rangle  \simeq  \epsilon  \,  (p\Delta^2)^{n/p -1} 
  \Gamma(n/p) \, .
\label{eq:scalmomentblanc}
\end{equation}
   In the following, we shall find the characteristics of the noise
 for which such an  anomalous  scaling is valid.

\subsection{The threshold shift}

 The presence of   noise can modify the bifurcation  threshold 
 which is given by $\epsilon = 0$ in the deterministic case.
 For the stochastic differential equation~(\ref{eq:Hopf}) 
 subject to an arbitrary noise $\xi(t)$, the critical value  
   $\epsilon_c(\Delta)$   that separates a localised PDF from 
 an  extended PDF,  is determined by  the vanishing
 of the  Lyapunov exponent associated with the fixed point
 $x = 0$. We linearize  equation~(\ref{eq:Hopf})
 around $x = 0$,
\begin{equation}
   \frac{ {\rm d}{ \delta{x}}  }{{\rm d}t} =
  \left( \epsilon   + \Delta \xi(t) \right) \delta{x} \, , 
\end{equation} 
 and find  the Lyapunov exponent $\Lambda$ to be   
 \begin{equation}
    \Lambda = \lim_{ t \to \infty} \frac{ {\rm d}}{{\rm d}t}
 \langle \ln( \delta{x} ) \rangle = 
  \langle \frac{  \frac{ {\rm d} {\delta{x}}  }{{\rm d}t}   }
 {\delta{x} } \rangle  =  \epsilon   \, . 
 \end{equation} 
 This result is valid regardless of the nature of the noise $\xi$
 (we have only used the fact that the mean value of $\xi$ vanishes).
 The  Lyapunov exponent vanishes when $\epsilon = 0$. Thus,
 for the first order equation~(\ref{eq:Hopf}), there is
 no threshold shift due to the presence of noise and we always  have
   $\epsilon_c(\Delta) = 0$.

\subsection{The Poincar\'e-Lindstedt expansion}

  In order to to simplify the general treatment,
  we shall study the case of a  cubic nonlinearity, {\it i.e.}, we
  take $p=1$ in equation~(\ref{eq:Hopf}).
   The Poincar\'e-Lindstedt method 
  \cite{drazin,kevorkian,luecke1,luecke2} consists in writing
  two   expansions~: one  for the solution $x(t)$ of 
  equation~(\ref{eq:Hopf}) and  another  for  the deviation
 from threshold  $(\epsilon - \epsilon_c(\Delta))$, both as 
  power  series of a formal parameter $\lambda$,   
\begin{eqnarray}
       x(t)  &=& \lambda  x_1(t) + \lambda^2 x_2(t) +  \lambda^3  x_3(t)
+  \lambda^4  x_4(t)  + \ldots       \, ,
 \label{PLE1}  \\ 
     \epsilon - \epsilon_c(\Delta)    &=&   \lambda
\epsilon_1 +  \lambda^2  \epsilon_2 + \lambda^3\epsilon_3 + 
 \lambda^4  \epsilon_4 + \ldots   \, \label{PLE2}   \, .
\end{eqnarray}
By substituting the expansions~(\ref{PLE1} and \ref{PLE2}) in 
 equation~(\ref{eq:Hopf}) we observe that $x(t)$ 
 and $(\epsilon - \epsilon_c(\Delta))$  are,  respectively, odd and 
 even  in  $\lambda$
 (because  equation~(\ref{eq:Hopf})  is antisymmetric under  $x \to -x$).
  The Poincar\'e-Lindstedt expansion thus  reduces to
  (using the fact that $  \epsilon_c(\Delta) = 0$)
 \begin{eqnarray}
       x(t)  &=& \lambda \left( x_1(t) + \lambda^2 x_3(t) +  \lambda^4  x_5(t)
    + \ldots     \right)      \, ,
 \label{PL1}  \\ 
     \epsilon    &=&   \lambda^2 \left( 
\epsilon_2 +  \lambda^2  \epsilon_4 + \lambda^4\epsilon_6 + 
   \ldots   \right)   \, \label{PL2}   \, .
\end{eqnarray}
 Substituting  these  formal expansions  in  equation~(\ref{eq:Hopf})
 and identifying  the terms  order by order in $\lambda$, we obtain 
   a hierarchy of first order  differential equations for the
  functions  $x_{2i+1}(t)$ 
\begin{equation}
  {\mathcal L}x_{2i+1}(t) =  {\mathcal P} \left( x_1, x_3, \ldots, 
 x_{2i-1}, \epsilon_2, \ldots, \epsilon_{2i} \right) \, ,
\label{eq:hierarchy}
\end{equation}
 where  ${\mathcal P}$ is a polynomial
  function and  $ {\mathcal L}$  a  linear differential
 operator~: 
\begin{equation}
   {\mathcal L}  = \frac{ {\rm d}}{ {\rm d}t} - \Delta \xi(t) \, . 
\label{eq:defL}
\end{equation}
  The hierarchy~(\ref{eq:hierarchy})  is  solved with
the initial conditions 
\begin{equation}
  x_1(0) = 1 \, , \hbox{ and }  \,\,\,\,\,\,  
 x_{2i +1}(0) = 0  \hbox{   for all }  i \ge 1  \, .
 \label{eq:init}
\end{equation}
 These  conditions imply that $x(0) = \lambda$~: 
 the formal parameter $\lambda$ is simply equal to the  value of $x$ 
 at time $t =0$ and,  therefore,  has the dimensions  
\begin{equation}
    \lambda  \sim t^{{1}/{2}}  \, .
\label{eq:dimlambda}
\end{equation}
 Besides,  the  numbers $\epsilon_i$ appear as parameters
 in equation~(\ref{eq:hierarchy}) and are determined 
  recursively thanks to the   solvability condition that we 
 now derive.  Let us call $y_1(t)$ the solution
 of the adjoint equation  
 ${\mathcal L}^ \dagger  y_1  = 0 $, which is  given by
\begin{equation}
  y_1(t)  =    \exp\left( -  \Delta  \int_0^t \xi(u){\rm d}u\right) \, .
\label{eq:defy1}
\end{equation}
 Multiplying   equation~(\ref{eq:Hopf}) by the function 
   $y_1$  and taking average values,  we obtain 
\begin{equation}
 \langle  y_1 {\mathcal L} x_1  \rangle  =
 \langle     \epsilon  y_1  x_1 -  y_1  x_1^3 \rangle    \, .
\end{equation}
 Integrating  the left hand side of this equation by parts
 and taking  into account the fact 
  that $y_1$ is in the kernel of the adjoint 
  operator  ${\mathcal L}^ \dagger$, we derive  the following
  relation 
\begin{equation}
   \epsilon  = \frac{  \langle  y_1  x_1^3 \rangle }  
 {  \langle  y_1  x_1\rangle }  \, .
\label{eq:solvability}
\end{equation}
  By virtue of  this solvability condition, 
 the hierachy of equations~(\ref{eq:hierarchy})
 is  defined without ambiguity. The functions $x_i$ and
 the parameters  $\epsilon_i$ are  determined  recursively 
 in a unique manner  using  the initial conditions~(\ref{eq:init}).

  Eliminating  $\lambda$
   from  equations~(\ref{PL1}) and~(\ref{PL2}) leads to 
 the  required  expansion of  $x(t)$ in terms of 
 $\epsilon$.

\subsection{Calculation of the first terms in the expansion}

   We now  calculate the first terms of the Poincar\'e-Lindstedt expansion
 by applying the procedure described above. 
  It will be useful to introduce  the following auxiliary  random variable
 $B_t$~:
 \begin{equation}
       B_t  =   \int_0^t \xi(u){\rm d}u \, . 
  \label{eq:defB}
\end{equation}
   Because  $\xi$ is  taken to be a  Gaussian random process,
  $B_t$ is also Gaussian. 
  The lowest order  terms  in  the Poincar\'e-Lindstedt expansion
 are then given by
\begin{eqnarray}
       x_1(t)  &=&    \exp\left( \Delta   B_t \right)  \label{eq:x1} \, ,\\
       x_3(t)  &=&  x_1(t)  \left( \epsilon_2 t -  
 \int_0^t x_1^2((u){\rm d}u  \right)  \, ,   
 \label{eq:x3} 
\end{eqnarray}
 and  the parameters $\epsilon_2$, $\epsilon_4$ are given by
\begin{eqnarray}
    \epsilon_2 &=&  \langle  x_1^2(t)  \rangle  =  
 \langle   \exp\left(  2 \Delta   B_t \right)   \rangle  \label{eq:e2} \, ,\\ 
   \epsilon_4  &=&  3 \langle  x_1x_3  \rangle  - 
    \epsilon_2   \Big\langle  \frac{x_3}{x_1}  \Big\rangle
  \label{eq:e4}    \, , 
\end{eqnarray}
 where  the expectation value $\langle . \rangle$ is taken over
 all the possible histories between times  0 and $t$.

\subsection{Behaviour of the moments}
 
   We  now  determine the  behaviour  of the moments of $x$, such as 
 $\langle x^{2n} \rangle $,  in the vicinity of the threshold, {\it i.e.},
  when $\epsilon \to 0$. In the Poincar\'e-Lindstedt expansion, this
  corresponds to   the formal parameter $\lambda$  converging   to 0.
 In this limit, we obtain
\begin{eqnarray}
 \langle x^{2n} \rangle &=& \lambda^{2n}
  \left( \langle x_1^{2n} \rangle  +  2n \lambda^2 
 \langle x_1^{2n-1} x_3 \rangle  + \ldots \right) 
  \simeq  \lambda^{2n} 
  \langle  \exp\left( 2n \Delta   B_t \right)\rangle 
  =  \lambda^{2n}  \exp\left( 2n^2 \Delta^2   \langle B_t^2 \rangle \right)  
     \, ,  \label{eq:nmom}\\
   \epsilon  &=& \lambda^2 
 \left(\epsilon_2 + \lambda \epsilon_4
   + \ldots \right) \simeq  \lambda^2
 \langle  \exp\left( 2\Delta   B_t \right)\rangle 
 =  \lambda^2 \exp\left( 2 \Delta^2   \langle B_t^2 \rangle \right) \, .
  \label{eq:epsmom}
\end{eqnarray}
 The last equality  is  derived  by using the fact that
 $B_t$ is a Gaussian random process. Eliminating $\lambda$
  from  equations~(\ref{eq:nmom}) and~(\ref{eq:epsmom}), we obtain 
\begin{equation}
   \langle x^{2n} \rangle \simeq  \epsilon^n 
   \exp\left( 2(n^2 -n) \Delta^2   \langle B_t^2 \rangle \right) \, .
\label{eq:scalmom}
\end{equation}
   This equation predicts a   normal 
  scaling behaviour identical to that  of  the deterministic case.
  In fact, this scaling
  is merely a   logical consequence of  the  formal  structure 
 of the Poincar\'e-Lindstedt expansion~:  indeed, we have
 $  \langle x^{2n} \rangle \sim \lambda^{2n}$
  and $\epsilon \sim \lambda^2$ and therefore 
 $  \langle x^{2n} \rangle \sim \epsilon^n$.  However, the 
  proportionality factor
 between  $ \langle x^{2n} \rangle $
  and $\epsilon^n $ 
 in  equation~(\ref{eq:scalmom})  can  be  divergent  when $t \to \infty$.
 Such a divergence can change  the  scaling behaviour of the moments
 in the large time limit.

\subsection{Importance of  low frequencies in  the noise spectrum}

   In order to determine  the multiplicative factor 
 of $\epsilon^n $  in equation~(\ref{eq:scalmom}), we must
 calculate the variance of the random variable $B_t$~:
\begin{eqnarray}
     \langle B_t^2 \rangle  =
   \int_0^t\int_0^t  \langle \xi(u) \xi(v) \rangle{\rm d}u {\rm d}v 
   =    \int_0^t\int_0^t   {\mathcal D}(|u-v|) \rangle{\rm d}u {\rm d}v 
   =   \int_{-\infty}^{+\infty} 
 \frac{ 1 - \cos\omega t}{\omega^2} \,\,
   \frac{ \hat{\mathcal D}(\omega) {\rm d}\omega}{\pi}\, .
  \label{eq:varB}
\end{eqnarray}
The last integral is well defined at $\omega  = 0$ (the time $t$
 introduces an effective low frequency  cut-off for $\omega \sim 1/t$).
 The behaviour of  $\langle B_t^2 \rangle$ for  $t \to \infty$
 depends on the  behaviour of  $\hat{\mathcal D}(\omega)$ at 
 $\omega  \to  0$. The following  two cases must be distinguished~:
 
 (i) The spectrum of the noise vanishes at 
  low frequencies, {\it i.e.},
 ${\mathcal D}(0) = 0$. Because  $\hat{\mathcal D}(\omega)$
 is an even function of $\omega$,  we suppose that
  $\hat{\mathcal D}(\omega)  \sim \omega^2 $ for
  $\omega \sim 0$ (we disgard  non-analytic
 behaviour of the power spectrum at the origin. Such   non-analyticity  would
 correspond to long tails in the correlation function of the noise). 

 (ii) The power spectrum of the noise is finite at  $\omega =  0$,
 {\it i.e.},  ${\mathcal D}(0) > 0$.

 In case (i), the long time limit of equation~(\ref{eq:varB}) 
 is readily derived and we obtain (by using the Riemann-Lebesgue
 lemma)
\begin{equation}
    \langle B_t^2 \rangle   \rightarrow  \int_{-\infty}^{+\infty} 
   \frac{ \hat{\mathcal D}(\omega) {\rm d}\omega}{ \pi \, \omega^2} \,\,\,\,\,
   \hbox{ when } \,\,\,\,\, t \to \infty \,. 
\end{equation}
The variance of $B_t$ has  a {\it finite} limit at large 
 times and, therefore,   the prefactor of
 $\epsilon^n$  in equation~(\ref{eq:scalmom}) converges to a finite
 number when $ t \to \infty$,  for all $ n \ge 1$.
  Thus,  the  Poincar\'e-Lindstedt expansion, used
 at the first order,  leads to  a well-defined asymptotic behaviour
 for $\langle x^{2n}\rangle $ and allows us to recover 
 the 'normal'  scaling behaviour   which was predicted in  \cite{luecke1}
 for the  random frequency  oscillator. 
 Higher order terms in the expansion do not affect  the scaling
 exponents  and modify  the prefactors only. The next order term was
 studied in  \cite{luecke2}.

 In case (ii), the integral on the right hand side of equation~(\ref{eq:varB})
 diverges when  $ t \to \infty$ and its leading behaviour is
\begin{equation}
 \langle B_t^2 \rangle  =  \frac{ t }{\pi } \int_{-\infty}^{+\infty} 
 \frac{ 1 - \cos u }{ u^2} \,\,
    \hat{\mathcal D}\left(\frac{u}{t}\right) {\rm d}u
   \rightarrow    \hat{\mathcal D}(0) t    \, .
\end{equation}
The variance of $B_t$  grows linearly with time in the long time
 limit. Thus,  for $n \ge 2$, the coefficient  of $\epsilon^n$ 
 in  equation~(\ref{eq:scalmom}) grows  exponentially with time.
  Such a  divergence of a  prefactor is an indication of anomalous scaling.
 This  anomalous scaling was not found in  \cite{luecke2}  where the
 authors  analyzed  only  the first few terms of the  
   Poincar\'e-Lindstedt expansion
  and supposed, in order to avoid divergencies,
 that the noise spectrum  has a low frequency 
 cut-off  $\omega_<$
 ({\it i.e.}, $\hat{\mathcal D}(\omega) = 0$ for $\omega \le \omega_<$). 
 We  will show in the next section that even 
   when low frequencies are present,
  the  Poincar\'e-Lindstedt expansion 
 can  be used  to extract sound results and  to  predict anomalous scaling.

\section{Resummation, Anomalous scaling and Intermittency}
\label{sec:resumation}

  When the power   spectrum   at $\omega = 0$
  is finite ({\it i.e.},  $ \hat{\mathcal D}(0) > 0$),
 all the coefficients of the
 Poincar\'e-Lindstedt series~(\ref{PL1},\ref{PL2})  blow up  
 when  time goes to infinity. This divergence, that  appears even 
  for  the lowest order,  seemingly  implies 
 that the expansion breaks down   when $t \to \infty$. 
  We  now show  that it is still  possible to use 
 the  Poincar\'e-Lindstedt expansion and  to extract from
 it by resummation   the multifractal behaviour  found in \cite{seb1}.

\subsection{Resummation of the  Poincar\'e-Lindstedt  series}

  Without loss  of generality, we    suppose in the sequel 
 that   $\hat{\mathcal D}(0) = 1$; this    amounts simply 
 to redefining  $\Delta$ as  $\Delta/\sqrt{\hat{\mathcal D}(0)}$.
 We first  analyse the lowest order terms  in  the    Poincar\'e-Lindstedt 
  expansion~(\ref{PL2},\ref{eq:e2}, \ref{eq:e4})
 of $\epsilon$. If  we retain  only  the most divergent  contribution
  when $ t \to \infty$,  we obtain
 \begin{equation}
    \epsilon_2 \simeq    \exp(2 \Delta^2 t)  \, , \,\,\,\,\,\, 
   \epsilon_4 \simeq   -  \frac{ \exp(8 \Delta^2 t) }{ 2\Delta^2 }
              \, , \,\,\,\,\,\, 
 \epsilon_6 \simeq  \frac{3}{8} \,  \frac{ \exp(18 \Delta^2 t) }
 { (2\Delta^2)^2 } \, . 
\label{devepsi}
 \end{equation}
 We now  investigate   the 
 general structure of  the  Poincar\'e-Lindstedt  s\'eries~(\ref{PL2}),
  retaining   for  each order   only  the most divergent term. 
  From  dimensional analysis (equations~(\ref{eq:dimensions})
 and~(\ref{eq:dimlambda})),   the dimensionless variable
 ${\epsilon}/{\Delta^2}$  can be written in terms of the dimensionless
 expansion parameter ${\lambda}/{\Delta}$ as follows 
\begin{equation}
   \frac{\epsilon}{\Delta^2}  = \sum_{k=1}^{\infty}
  (-1)^{k-1} a_k \left( \frac{\lambda^2}{\Delta^2}\right)^{k}
          \exp( 2k^2 \Delta^2 t )\, ; 
\label{eq:serieseps}
\end{equation}
  using equation~(\ref{devepsi}), we  find 
 $a_1 = 1, a_2 = 1/2 , a_3 = 3/32$. In fact, the general
 formula for the $k$-th coefficient $a_k$ can be obtained   recursively
  using  equation~(\ref{eq:hierarchy})
\begin{equation}
   a_k = \frac{k}{ 4^{k-1} (k-1)!}    \,. 
\label{eq:formak}
\end{equation}
  We emphasize that the series~(\ref{eq:serieseps}) is 
  divergent: its  radius of convergence is strictly 0. 
 However,  it is possible to  make a resummation  of  this  divergent 
  expansion  
  by using the following  representation  for  the exponential term
\begin{equation}
     \exp( k^2)  =   \int_{-\infty}^{+\infty} \frac{{\rm d}u }{\sqrt{\pi}}
       \exp( -u^2 + 2ku)                \, .
\label{eq:sommation}
\end{equation}
Inserting this identity in  equation~(\ref{eq:serieseps}), we obtain
\begin{equation}
   \frac{\epsilon}{\Delta^2}  = \int_{-\infty}^{+\infty}
   \frac{{\rm d}u }{\sqrt{2\pi}} \exp\left( - \frac{u^2}{2}\right)  
   \sum_{k=1}^{\infty}
  (-1)^{k-1} a_k \left( \frac{\lambda^2}{\Delta^2}\right)^{k}
          \exp( 2k \Delta \sqrt{ t} u )  \, . 
\label{eq:ressumingeps}
\end{equation}
  The series under the integral sign has  generically
 a non-zero  radius of convergence that depends on the
 coefficients  $a_k$. Defining  
\begin{equation}
 F(z) =  \sum_{k=1}^{\infty}  (-1)^{k-1} a_k z^k \, ,
\label{eq:defF}
\end{equation}
 and  substituting   in this equation the expression~(\ref{eq:formak})
  for the $a_k$'s, we obtain
 \begin{equation}
  F(4z) = 4z( 1 - z)\exp(-z) \, .
\label{eq:Fexplicit}
 \end{equation}
  Equation~(\ref{eq:ressumingeps}) can now be written  as follows
\begin{equation}
   \frac{\epsilon}{\Delta^2}  =    \int_{-\infty}^{+\infty} 
 \frac{{\rm d}u }{\sqrt{2\pi}} \exp\left( - \frac{u^2}{2}\right) 
     F\left(\frac{\lambda^2}{\Delta^2} 
    \exp(2 \Delta \sqrt{ t} u )     \right)  \, .
\label{eq:ressumingeps2}
\end{equation}

 A  resummation   can be performed  along the same lines for
 the mean value of   $x^{2n}$~: starting from the
   Poincar\'e-Lindstedt  expansion~(\ref{PL1}), we find that 
\begin{equation}
   \frac{\langle x^{2n} \rangle}{\Delta^{2n}}  = \sum_{k=0}^{\infty}
  (-1)^{k} b_k^{(n)} \left( \frac{\lambda}{\Delta}\right)^{2(n+k)}
          \exp( 2(n+k)^2 \Delta^2 t )\, .
\label{eq:seriesmom}
\end{equation}
 Using  equation~(\ref{eq:sommation})  this divergent series is transformed
 as  
\begin{equation}
  \frac{\langle x^{2n} \rangle}{\Delta^{2n}}   =    \int_{-\infty}^{+\infty} 
 \frac{{\rm d}u }{\sqrt{2\pi}} \exp\left( - \frac{u^2}{2}\right) 
    G_n\left(\frac{\lambda^2}{\Delta^2} 
    \exp(2 \Delta \sqrt{ t} u )     \right) \, , 
\label{eq:ressumingmom}
\end{equation}
where  we have introduced the new function 
\begin{equation}
G_n(z) =   \sum_{k=0}^{\infty}  (-1)^{k} b_k^{(n)}  z^{n+k}   \, .
 \end{equation}
  
 To summarize, we have resummed the Poincar\'e-Lindstedt series
 and obtained the following relations~:
\begin{eqnarray}
    \frac{\epsilon}{\Delta^2}  &=
   {\mathcal F}\left(\frac{\lambda^2}{\Delta^2}, \Delta \sqrt{ t} \right) 
       \,\,\, &\hbox{ where } \,\,\,
   {\mathcal F}(z, w) \sim z  \,\,\,  \,\,\, \hbox{  when  } \,\,\,
  z \to 0 \hbox{  and    $w$ is finite}
      \, , \label{ResumPLeps} \\
       \frac{\langle x^{2n} \rangle}{\Delta^{2n}}    &=
    {\mathcal G}_n\left(\frac{\lambda^2}{\Delta^2}, \Delta \sqrt{ t} \right)
      \,\,\, &\hbox{ where } \,\,\,
   {\mathcal G}_n(z, w) \sim z^n  \,\,  \hbox{ when  } \,
  z \to 0 \hbox{  and $w$ is finite}
   \, .  \label{ResumPLmom}            
 \end{eqnarray}
 The scaling behaviour of $\langle x^{2n} \rangle$ as a function
 of $\epsilon$ in the vicinity
 of the bifurcation threshold   is obtained by eliminating $\lambda$ between
  these two equations.

  \subsection{Anomalous scaling}

The functions  ${\mathcal F}$ and
  ${\mathcal G}_n$  that appear in equations~(\ref{ResumPLeps} and
  \ref{ResumPLmom}) have a singular behaviour when 
 $\lambda \to 0$ and $ t \to \infty$~:
 these two limits {\it  do not commute}.
   However, thanks to
   the expressions given in equations~(\ref{ResumPLeps}, \ref{ResumPLmom}), 
   we can  disentangle these two limits.

 If we take  $\lambda \to 0$    for a large but  fixed  value of $t$,
 we find  from equations~(\ref{ResumPLeps}) 
  and~(\ref{ResumPLmom}) that $\langle x^{2n} \rangle \sim \epsilon^n$.
 Normal scaling is therefore  recovered,  in agreement with 
 equation~(\ref{eq:scalmom}).

  However, taking  the 
 limit $ t \to \infty$ first and keeping   the value of  $\lambda$ 
 fixed and small, we obtain from  equation~(\ref{eq:ressumingeps2}), 
\begin{equation}
   \frac{\epsilon}{\Delta^2}  =  
  \int_{-\infty}^{+\infty} \frac{{\rm d}v}{2 \Delta \sqrt{2\pi t} } \,
 \exp\left( - \frac{v^2}{8\Delta^2 t  }\right) 
     F\left(\frac{\lambda^2}{\Delta^2}  \exp(v ) \right) 
 \simeq 
  \int_{-\infty}^{+\infty} \frac{ {\rm d}v }{2 \Delta \sqrt{2\pi t} } 
   F\left(\frac{\lambda^2}{\Delta^2}  \exp(v ) \right)  
 =   \frac{\int_{-\infty}^{+\infty}  {\rm d}v \,F\left(\exp(v ) \right)}
     {2 \Delta \sqrt{2\pi t} }       \, , 
\label{eq:limeps}
\end{equation}
  where the last equality is obtained by  the translation 
 $ v \to v - \log({\lambda^2}/{\Delta^2})$.
    Using the explicit expression~(\ref{eq:Fexplicit}) for 
 $F(z)$ we  verify that the integral on  the right-hand
 side of equation~(\ref{eq:limeps}) is a strictly positive  real number, 
 {\it i.e.,} it is neither zero nor infinite. We have thus shown that
 \begin{equation}
   \frac{\epsilon}{\Delta^2} \simeq  \frac{ e_0 }{ \Delta \sqrt{2\pi t} } 
 \, ,  
\label{eq:limeps2}
\end{equation}
 where  $e_0 > 0\,.$
  In a similar manner,  using equation~(\ref{eq:ressumingmom}),
 we  find  that the asymptotic behaviour of the $2n$-th moment of
 $x$ is given by
    \begin{equation}
   \frac{\langle x^{2n} \rangle}{\Delta^{2n}} 
  \simeq  \frac{ c_n }{ \Delta \sqrt{2\pi t} }  \, . 
\label{eq:limmom}
\end{equation}
 Eliminating $t$ from  equations~(\ref{eq:limeps2}) and~(\ref{eq:limmom})
   provides us the scaling behaviour of the moments of $x$
  in the vicinity of the bifurcation threshold
  \begin{equation}
  {\langle x^{2n} \rangle}   \simeq C_n 
  \epsilon  \Delta^{2n -2}   \, . 
\label{eq:anomscaling}
\end{equation}
  In the vicinity of  $ \epsilon = 0$ all the moments scale  
  linearly  with $\epsilon$. This equation 
   generalizes  the calculation of
  the white noise case~(\ref{eq:scalmomentblanc})
 to an {\it arbitrary } noise with non-vanishing 
 zero-frequency power spectrum. 
 We have thus shown, using  the Poincar\'e-Lindstedt expansion, 
 that the low frequency of the noise strongly  affect
  the scaling  of the  order parameter 
 in the vicinity of the threshold of a stochastic  bifurcation
  and induces a multifractal behaviour; a qualitative  explanation  of 
 this effect was given  in \cite{seb1}.

\subsection{Discussion of the random frequency oscillator}
 \label{sec:model}

 The  above analysis   can be applied to more general
 random dynamical systems such as the  
   parametrically driven damped anharmonic oscillator  that  
 naturally appears in the study of many instabilities \cite{fauve}.  
 Such a system   is described  by the  following equation~:
\begin{equation}
   m  \ddot{x} + m \gamma \dot{x} =
  \left(\epsilon + \Delta \xi(t) \right) x -  x^3 \, ,  
 \label{eq:Lucke1}
\end{equation}
  where  $\epsilon$ is  the control parameter  and  the modulation
$\xi(t)$  is of arbitrary dynamics and statistics: it can be a
periodic function or  a  random noise.
   For small driving amplitudes
 $\Delta$,  L\"ucke and Schank   \cite{luecke1} have 
  performed   a Poincar\'e-Lindstedt expansion, and obtained  an expression 
 for the threshold   $\epsilon_c(\Delta)$  (at first order in
 $\Delta$). Their result 
 has been verified  both  numerically and 
 experimentally and is  also in agreement with
 the exact  result obtained for the  Gaussian white noise
 (in this case a closed formula is available for
 $\epsilon_c(\Delta)$   for  arbitrary values
 of $\Delta$).  Another result  obtained in   \cite{luecke1,luecke2}
 is the scaling of the moments near the threshold, 
\begin{equation}
  \langle x^{2n} \rangle = s_n 
\left( \epsilon - \epsilon_c(\Delta) \right)^{n} 
 + {\mathcal O}\left( ( \epsilon - \epsilon_c)^{n + 1}  \right)    \, ,
 \label{eq:Luckemoment}
\end{equation}
 where the constant $s_n$ depends on  $\xi(t)$ and on $\Delta$. The
 moments have   a  {\it normal scaling}  behaviour~:
 $ \langle x^{2n} \rangle$ scales as  $\langle x^2 \rangle^n$.   
 The bifurcation scaling exponent is equal to  1/2  and is
   the same as that of  a   deterministic Hopf bifurcation.  However,  
  this expression does not agree with the 
    results for   random iterated maps,    for 
  the random parametric  oscillator  and  
   with  recent  studies  on  On-Off intermittency 
   \cite{pikovsky, philkir1,seb1}.  These works 
  predict that   the variable   $x$  is   intermittent  and  that 
  the   moments of    $x$  exhibit  {\it  anomalous scaling}, 
\begin{equation}
   \langle x^{2n} \rangle  \simeq \kappa_n  (\epsilon - \epsilon_c) 
      \,\,\,  \hbox{  for all } \,\,\, n >0 \, ,
 \label{eq:anomalous}
\end{equation}
 {\it i.e.},  all the moments  grow linearly 
 with the distance from  threshold. This multifractal behaviour,
 confirmed by numerical simulations for a  Gaussian white noise,  
  was  derived using  effective  Fokker-Planck equations.

 The origin of the  contradiction  between
  equations~(\ref{eq:Luckemoment})
  and (\ref{eq:anomalous})  lies in  the  divergences
 that appear in the Poincar\'e-Linsdtedt expansion. This fact,  
  identified  in  \cite{luecke2},  implies that   the results of
  \cite{luecke1} are valid only for  noises 
  that do  not have  low frequencies. We have shown  by studying 
 a model  technically simpler 
   than  equation~(\ref{eq:Lucke1}),  that 
  the   perturbative  Poincar\'e-Lindstedt expansion 
 can be used for any kind of noise by 
  resumming the divergent terms  to all orders. This resummation
   allows to recover  the  scaling exponents for  all the cases, 
  and  describes  the crossover between  the scalings
  given in  equations~(\ref{eq:Luckemoment})
  and~(\ref{eq:anomalous}).

 \section{Conclusion}

   In this work, we have shown that  the  Poincar\'e-Lindstedt  expansion,
 a  classical  perturbative technique 
 extensively used in the field of nonlinear dynamical systems, 
 can be successfully adapted  to analyze   a stochastic  model 
 that  plays the role of a paradigm for noise-induced bifurcations. 
 These perturbative expansions for stochastic dynamics
  were studied   in \cite{luecke1}, but,  due to the presence of 
 divergent terms,    were  
   applied  only   to  random noises  with  vanishing  power spectrum 
 at low frequencies (e.g.,  noises with a low frequency cut-off)
  \cite{luecke2}. However,  we have  shown 
 here  that,   by  a
  resummation of  the  divergent terms, 
   the Poincar\'e-Lindstedt method 
 can be  used for   {\it any type of noise}.
 Moreover,   the divergences that appear in the   Poincar\'e-Lindstedt
 expansion  are not  mathematical artefacts  but   have  genuine
 physical consequences~: the presence of low frequencies
 in the noise spectrum (that  leads to 
 these   divergences)   modifies the scaling behaviour of the
 order parameter in the vicinity of the bifurcation  threshold.
 If low frequencies are absent from the noise spectrum,
  the  order parameter has the same scaling as that of
  a deterministic bifurcation. In contrast, if the power spectrum
 does not vanish at zero frequency, the  order parameter exhibits
 anomalous  scaling in agreement with recent results for
 white and harmonic noise.

  Our work  has allowed us to analyze precisely the role of low frequencies
 of the noise in  a first order random dynamical system.  
 The  resummation  technique  we have used is fairly general 
  and we  believe that  the  results   we have derived 
  remain  valid for  systems of  higher order   in the
   vicinity of  a stochastic Hopf bifurcation.

\end{document}